\def\be{\begin{equation}}
\def\ee{\end{equation}}
\def\ba{\begin{array}}
\def\ea{\end{array}}

\newcommand{\tr}{\mathrm{tr}}

\documentclass[aps,pra,preprint,showpacs,amsmath,amssymb,amsfonts,preprintnumbers]
{revtex4}
\usepackage{epsfig,graphicx, color}
\usepackage{amsmath}

\newtheorem{theorem}{Theorem}
\newtheorem{lemma}{Lemma}
\begin{document}

\baselineskip=22pt \setcounter{page}{1} \centerline{\Large\bf
An Upper Bound of Fully Entangled Fraction of Mixed States} \vspace{6ex}
\begin{center}
 Xiaofen Huang$^{1}$,  Naihuan Jing$^{2,3}$, Tinggui Zhang$^{1\ast}$
\bigskip

\begin{minipage}{6.5in}
$^1$ School of Mathematics and Statistics, Hainan Normal University, Haikou, 571158, China\\
$^2$ School of Mathematics, South China University of
Technology, Guangzhou, 510640, China\\
$^3$ Department of Mathematics, North Carolina State University,
Raleigh, NC27695, USA\\
$\ast$Corresponding author: tinggui333@163.com

\end{minipage}
\end{center}
\vskip 1 true cm
\begin{center}
\begin{minipage}{5in}
\vspace{3ex} \centerline{\large Abstract} \vspace{4ex}

We study the fully entangled fraction of a quantum state. An upper bound is obtained for arbitrary bipartite system. This upper bound only depends on the Frobenius norm of the state.
\end{minipage}
\end{center}



The fully entangled fraction is closely related to many quantum information processings, such as quantum computation \cite{inf}, quantum teleportation \cite{telep}, dense coding \cite{code}, quantum cryptographic schemes \cite{cryp}, entanglement swapping \cite{swap}, and remote state preparation (RSP) \cite{RSP1, RSP2, Liu2, RSP3} etc. For instance, in the process of teleportation, the fidelity of optimal teleportation is given by fully entangled fraction (FEF) \cite{FEF}. Thus an analytic formula for FEF is of great
importance. In \cite{Lag} an elegant formula for a two-qubit system is derived analytically by using
the method of Lagrange multipliers. Concerning the estimation of entanglement of formation and concurrence, exact results have been obtained not only for two-qubit case, but also
for some higher dimensional states, isotropic and Werner states \cite{ZMJ}. Analytical lower bounds
have also been obtained for general cases \cite{lower1, lower2}. In \cite{LM} an estimate of the upper bound of FEF was given. Some relations between FEF with eigenvalues of the density matrix were studied in \cite{GRJ}. Nevertheless, analytical computation of FEF remains formidable and few results have been known for higher dimensional quantum states.

The aim of this work is to give an upper bound of the FEF for arbitrary higher dimensional state. Our main techniques come from
a careful analysis of the Frobenious norm.

\section{Introduction}
Consider the bipartite quantum state $\rho$ in Hilbert space $\mathcal{H}\otimes \mathcal{H}$£¬ where $\mathcal H$ has  the computational basis $|i\rangle$, $i=1,2, \ldots, d$. Let $E_{ij}$ be the $d\times d$ unit matrix with the only nonzero entry 1 at the position $(i, j)$.
Let $\omega$ be a fixed $d$th primitive root of unity. Consider the {\it principal basis} matrices
%{\bf{Definition} }The principal matrices are with the form
\begin{equation}\label{principal matrix}
    A_{ij}=\sum_{m\in Z_d}\omega^{im}E_{m, m+j},
\end{equation}
where $\omega^d=1$, $i,j\in Z_d$, and $Z_d$ is $Z$ modulo $d$.

It is well-known that $\{A_{ij}\}$ is a set of linear generators of the general linear Lie algebra $\mathfrak{gl}(d)$. In the case of $d=2$, the principal matrices specialize to the Pauli matrices, but in general they are different from the Cartan-Weyl basis and Gell-Mann basis. This
does not decrease its importance in quantum algebra. For instance, they play an important role in affine Lie algebras and
Yangians (cf. \cite{Liu}). The set $\{A_{ij}\}$ spans the principal Cartan subalgebra of $\mathfrak{gl}(d)$. Under the stand bilinear form $(x | y)=\tr(xy)$, the dual basis of the principal basis $\{A_{ij}\}$ is $\{\frac{\omega^{ij}}{d}A_{-i,-j}\}$. This also follows from the algebraic property
of the principal matrices
$$
A_{ij}A_{kl}=\omega^{jk}A_{i+k,j+l}.
$$
Then $A_{i, j}^{\dagger}=\omega^{ij}A_{-i, -j}$, so $\tr(A_{ij}A_{kl}^{\dagger})=\delta_{ik}\delta_{jl}d$.

Now we fix $j\in Z_d$, and define the Toeplitz sequence $\{\epsilon_j\}$ by $\epsilon_j(k)=E_{k,k+j}$, where $k$ is the index variable. Then the principal basis elements $A_{ij}$ can be written as the discrete Fourier transform:
$$
A_{ij}=F(\{\epsilon_j\})(i)=\sum_{m\in Z_d}\omega^{im}E_{m,m+j}.
$$
Therefore the usual Cartan-Weyl basis can be easily computed by the inverse Fourier transform
$$
E_{k, k+j}=\epsilon_j(k)=\frac{1}{d}\sum^{d-1}_{l=0}\omega^{-kl}A_{lj}.
$$

It is straightforward to get the trace of $A_{ij}$,
$$
tr(A_{ij})= \left\{
                         \begin{array}{ll}
                           0, & i\neq 0 ~or~ j\neq0; \\
                           d, & i=j=0.
                         \end{array}
                       \right.
$$

The Gell-Mann basis of $su(d)$ can be defined as the set of unitary matrices $\{\lambda_i\}$ of size $d\times d$, $i=1,..., d^2-1$ with the orthogonality relation $Tr(\lambda_i\lambda_j)=2\delta_{ij}$, which are used in the Bloch representation \cite{Bloch}. The matrices $\{\lambda_i\}$ can be constructed by another orthogonal basis $\{|a\rangle\}_{a=0}^{d-1}$ in space $\mathcal{H}$ \cite{cons}. Let $i, j, k$ be the indices such that $0\leqslant l\leqslant d-2$ and $0\leqslant j<k\leqslant d-1$. Then for $i=1, ..., d-1$,
$$
\lambda_i=\sqrt{\frac{2}{(l+1)(l+2)}}(\sum_{a=0}^{l}|a\rangle\langle a|-(l+1)|l+1\rangle\langle l+1|),
$$
and for $i=d, ... , \frac{(d+2)(d-1)}{2}$,
$$
\lambda_i=|j\rangle\langle k|+|k \rangle\langle j|,
$$
and for $i=\frac{d(d+1)}{2}, ... , d^2-1$,
$$
\lambda_i=-i(|j\rangle\langle k|-|k\rangle\langle j|).
$$
Obviously, the principal matrices have much simpler representation than $\{\lambda_i\}$. We will take advantage of their relative easy form.

\section{Upper Bound of Fully Entangled Fraction }
The fully entangled fraction of a density matrix $\rho$ is defined by
\begin{equation}\label{FEF}
 {\mathcal{F}(\rho)}=\mathop{max}_{U}\langle\varphi_{+}|(I\otimes U^{+})\rho (I\otimes U)|\varphi_{+}\rangle,
\end{equation}
where $|\varphi_{+}\rangle=\frac{1}{\sqrt{d}}\sum_{i=1}^{d}|ii\rangle$ is the maximal entangled state.

Let us represent $\rho$ in terms of the principal basis matrices:
\begin{equation}\label{REP}
 \rho=\frac{1}{d^2}[I\otimes I+\sum_{(i, j)\neq (0,0)}a_{ij}A_{ij}\otimes I+\sum_{(i, j)\neq (0, 0)} b_{ij}I\otimes A_{ij}+\sum_{(i, j), (k, l)\neq (0, 0)}c_{ij}^{kl}A_{ij}\otimes A_{kl}],
\end{equation}
where the coefficients
\begin{equation}\label{a[ij]}
a_{ij}=\omega^{ij}tr((A_{-i, -j}\otimes I)\rho)=tr(\rho(A_{ij}^{\dagger}\otimes I)),
\end{equation}
\begin{equation}\label{b[ij]}
b_{ij}=\omega^{ij}tr((I \otimes A_{-i, -j})\rho)=tr(\rho(I\otimes A_{ij}^{\dagger})),
\end{equation}
\begin{equation}\label{c[ijkl]}
c_{ij}^{kl}=\omega^{ij+kl}tr((A_{-i, -j} \otimes A_{-k, -l})\rho)=tr(\rho(A_{ij}^{\dagger}\otimes A_{kl}^{\dagger})).
\end{equation}

\textbf{ Example 1:}
Isotropic state \cite{iso}
$\rho_{iso}=\frac{1-p}{d^2}I\otimes I+p|\varphi_{+}\rangle\langle\varphi_{+}|$.  According to the decomposition by principal matrices, we have
\begin{equation}
   \rho_{iso}=\frac{1}{d^2}(I\otimes I +\sum_{(0, 0)\neq (0, 0)} pA_{ij}\otimes A_{-i, -j}).
\end{equation}

\textbf{Example 2.}
Werner entangled state \cite{wer}
$\rho_{w}=\frac{d+1}{d^3}I\otimes I-\frac{1}{d^2}P$,
where $P$ is the flip operator $P=\sum_{i}|i\rangle\langle j|\otimes|i\rangle\langle j|$. The Werner state has the following representation in terms of the principal matrices,
$$
\rho_{w}=\frac{1}{d^2}(I\otimes I-\frac{1}{d}\sum_{(0, 0)\neq (0, 0)} A_{ij}\otimes A_{-i,-j}).
$$
So by the principal presentation, the Werner state $\rho_{w}$ is actually a special isotropic state with $p=-d$!

The above two examples depend on the following result.
\begin{lemma} We have that \be\label{mix decomp}
  |\varphi_+\rangle\langle \varphi_+| = \frac{1}{d^2}I\otimes I+\sum_{(ij)\neq (00)}\frac{1}{d^2}A_{ij}\otimes A_{-i,-j}.
\ee \end{lemma}
 {\bf{Proof:}} We represent
$|\varphi_+\rangle\langle \varphi_+|$ in terms of the principal basis elements:
\begin{equation}\label{Rep}
  |\varphi_+\rangle\langle \varphi_+|=\frac{1}{d^2}[I\otimes I+\sum_{(i, j)\neq (0, 0)}a_{ij}A_{ij}\otimes I+
\sum_{(i, j)\neq (0, 0)}b_{ij}I\otimes A_{ij}+\sum_{(i, j), (k, l)\neq (0, 0)}c_{ij}^{kl}A_{ij}\otimes A_{kl}].
\end{equation}

Since
%\begin{eqnarray*}
%% \nonumber to remove numbering (before each equation)
% \langle\varphi_{+}|(I\otimes U^{+})\rho (I\otimes U)|\varphi_{+}\rangle &=& tr(\langle\varphi_{+}|(I\otimes U^{+})\rho (I\otimes U)|\varphi_{+}\rangle) \\
%  &=& tr(\rho (I\otimes U)|\varphi_{+}\rangle\langle\varphi_{+}|(I\otimes U^{+})),
%\end{eqnarray*}
\begin{equation}\label{phi}
|\varphi_{+}\rangle\langle\varphi_{+}|=\frac{1}{d}\sum_{i, j}|ii\rangle\langle jj|=\frac{1}{d}\sum_{i,  j}E_{ii, jj}=\frac{1}{d}\sum_{i, j}E_{ij}\otimes E_{ij}.
\end{equation}
Computing the coefficients, we have
\begin{eqnarray*}
% \nonumber to remove numbering (before each equation)
  a_{ij} &=& \omega^{ij}tr[(A_{-i,-j}\otimes I)|\varphi_+\rangle\langle \varphi_+|] \\
   &=& \omega^{ij}tr (\sum_{m}\omega^{-im}E_{m, m-j}\otimes I)(\sum_{k, l}\frac{1}{d}E_{kl}\otimes E_{kl})  \\
   &=&\frac{\omega^{ij}}{d}tr(\sum_{m, k, l}\omega^{-im}\delta_{m-j, k}E_{ml}\otimes E_{kl}).
\end{eqnarray*}
When $k=l=m$, $j=0$, we get ($i\neq 0$)
\begin{eqnarray*}
a_{ij} &=&\frac{\omega^{ij}}{d}tr(\sum_m\omega^{-im}\delta_{m-j, m}E_{mm}\otimes E_{mm})\\
       &=&\frac{\omega^{ij}}{d}\sum_m\omega^{-im}\delta_{m-j, m}=\frac{\omega^{ij}}{d}\sum_m\omega^{-im}
=0.
\end{eqnarray*}
So $a_{ij}=0=b_{ij}$.

Also,
\begin{eqnarray*}
% \nonumber to remove numbering (before each equation)
  c_{ij}^{kl} &=& \omega^{ij+kl}tr[(A_{-i, -j}\otimes A_{-k, -l})|\varphi_+\rangle\langle \varphi_+|] \\
   &=&\frac{\omega^{ij+kl}}{d}tr [\sum_{m,m^{\prime}, s, t}\omega^{(-im-km^{\prime})}(E_{m, m-j}\otimes E_{m^{\prime}, m^{\prime}-l})(E_{st}\otimes E_{st})] \\
   &=& \frac{\omega^{ij+kl}}{d}tr [\sum_{m, m^{\prime}, s, t}\omega^{(-im-km^{\prime})}\delta_{m-j, s}\delta_{m^{\prime}-l, s}E_{m, t}\otimes E_{m^{\prime}, t}].
\end{eqnarray*}
When $m^{\prime}=m=t$, $t-j=t-l$, we get
\begin{eqnarray*}
c_{ij}^{kl} &=& \frac{\omega^{ij+kl}}{d}\sum_{s, t}\omega^{(-i-k)t}\delta_{t-j, s}\delta_{t-l, s}
=\frac{\omega^{ij+kl}}{d}\sum_{s, t}\omega^{(-i-k)t}\delta_{t-j, s}\\
 &=& \frac{\omega^{ij+kj}}{d}\sum_{t-s=j}\omega^{(-i-k)t}
 =\frac{\omega^{ij+kj}}{d}\sum_{s=0}^{d-1}\omega^{(-i-k)(j+s)}\\
 &=& \frac{\omega^{ij+kj}}{d}\omega^{(-i-k)j}\sum_{s=0}^{d-1}\omega^{(-i-k)s}
=1.
\end{eqnarray*}
Thus, the equation (\ref{mix decomp}) holds.

\begin{theorem} If
$\rho$ is a bipartite state on the space
${\mathcal{H}}\otimes {\mathcal{H}}$, then the fully entangled
fraction of $\rho$ satisfies the following relation
\begin{equation}\label{bound}
 {\mathcal{F}(\rho)}\leqslant \frac{1}{d^2}+\frac{d-1}{d}\parallel \rho \parallel_{F},
\end{equation}
where $\parallel\rho\parallel_F=(\tr\rho\rho^{\dagger})^{1/2}$ is the Frobenius norm.
\end{theorem}

\textbf{Proof} It follows from definition that \begin{eqnarray*}
% \nonumber to remove numbering (before each equation)
 {\mathcal{F}(\rho)} &=& \mathop{max}_{U}tr[\rho(I\otimes U) |\varphi_+\rangle\langle \varphi_+|(I\otimes U^{+})] \\
   &=& \mathop{max}_{U}tr[\frac{1}{d^2}\rho(I\otimes U)(I\otimes U^{+})+\sum_{(ij)\neq (00)}\frac{1}{d^2}\rho(I\otimes U)A_{ij}\otimes A_{-i,-j}(I\otimes U^{+})] \\
   &=& \frac{1}{d^2}+\mathop{max}_{U}\sum_{(ij)\neq (00)}\frac{1}{d^2}tr[\rho(A_{ij}\otimes UA_{-i, -j}U^{+})].
\end{eqnarray*}

Since $tr[\rho(A_{ij}\otimes UA_{-i, -j}U^{+})]=\langle\rho,
A_{ij}\otimes UA_{-i, -j}U^{+}\rangle$, by H\"{o}lder inequality,
 $$\mid\langle\rho, A_{ij}\otimes UA_{-i, -j}U^{+}\rangle\mid\leqslant \parallel \rho \parallel_{F}\parallel A_{ij}\otimes UA_{-i, -j}U^{+}\parallel_{F}.$$

where the norm is Frobenius norm, i.e. $\parallel A \parallel_{F}=(trA^{\dagger} A)^{\frac{1}{2}}$, in the case of $i=0$,
$\parallel A_{ij}\parallel_{F}=\parallel A_{-i, -j}\parallel_{F}=\sqrt{d}$, otherwise,$\parallel A_{ij}\parallel_{F}=0$. Because the Frobenius norm is invariant under the unitary matrix, we have the upper bound in the Theorem.

The upper bound derived in \cite{LM} says that for any $\rho \in \mathcal{H}\otimes \mathcal{H}$, the fully entangled fraction $\mathcal{F}(\rho)$ satisfies
$$
{\mathcal{F}(\rho)}\leqslant \frac{1}{d^2}+4\parallel M^{T}(\rho)M(P_{+})\parallel_{KF},
$$
where $M(\rho)$ denotes the correlation matrix with the entries $m_{ij}$ given in the Bloch representation of $\rho$:
$$
\rho=\frac{1}{d^2}I\otimes I+\frac{1}{d}\sum_{i=1}^{d^2-1}r_i(\rho)\lambda_i\otimes I+\frac{1}{d}\sum_{j=1}^{d^2-1}s_j(\rho)I\otimes \lambda_j+\sum_{i, j=1}^{d^2-1}m_{ij}(\rho)\lambda_i \otimes \lambda_j,
$$
where $r_i=\frac{1}{2}tr\{\rho\lambda_i\otimes I\}$,
$s_j=\frac{1}{2}tr\{\rho I\otimes \lambda_j\}$,
$m_{ij}=\frac{1}{4}tr\{\rho\lambda_i \otimes \lambda_j\}$, $P_{+}$
stands for the projection operator to $|\varphi_{+}\rangle$, $M^{T}$
stands for the transpose of $M$, $\parallel
M\parallel=tr\sqrt{MM^{+}}$ is the Ky Fan norm of $M$.

Because of the complexity of $\{\lambda_i\}$, it is hard to compute the upper bound for a general state. There is an another upper bound of $\mathcal{F}(\rho)$ given in \cite{GRJ}, which is related to the eigenvalues.

\noindent {\bf Example 3:} We consider the bound entangled
state \cite{exm2}
$$\rho(a)=\frac{1}{8a+1}\left(
  \begin{array}{ccccccccc}
    a & 0 & 0 & 0 & a & 0 & 0 & 0 & a \\
    0 & a & 0 & 0 & 0 & 0 & 0 & 0 & 0 \\
    0 & 0 & a & 0 & 0 & 0 & 0 & 0 & 0 \\
    0 & 0 & 0 & a & 0 & 0 & 0 & 0 & 0 \\
    a & 0 & 0 & 0 & a & 0 & 0 & 0 & a \\
    0 & 0 & 0 & 0 & 0 & a & 0 & 0 & 0 \\
    0 & 0 & 0 & 0 & 0 & 0 & \frac{1+a}{2} & 0 & \frac{\sqrt{1-a^2}}{2} \\
    0 & 0 & 0 & 0 & 0 & 0 & 0 & a & 0 \\
    a & 0 & 0 & 0 & a & 0 & \frac{\sqrt{1-a^2}}{2} & 0 & \frac{1+a}{2} \\
  \end{array}
\right).
$$

\begin{figure}
\begin{center}
{\includegraphics{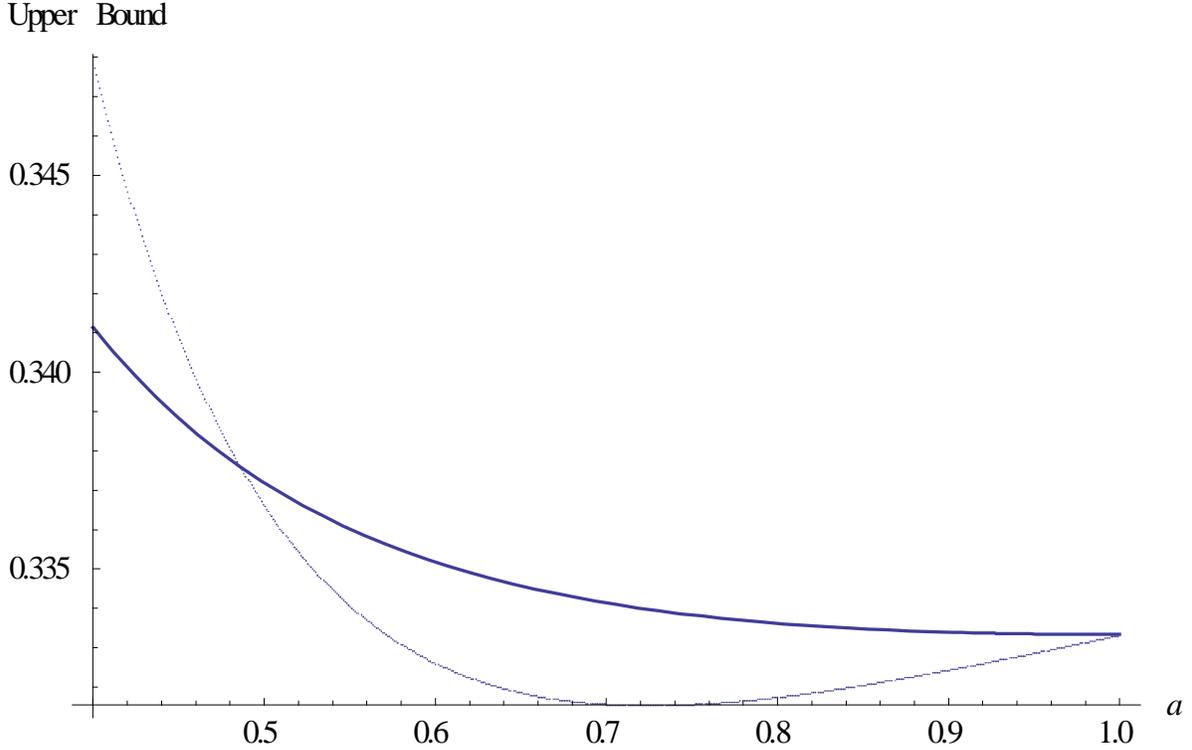}}
\end{center}
\caption{Our upper bound of ${\mathcal{F}\rho}$ from
($\ref{bound}$) (solid line) and the upper bound in \cite{GRJ} (dashed
line)}\label{Fig.1}
\end{figure}

We can compute the upper bound of the fully entangled fraction by
Theorem 1,
${\mathcal{F}\rho}\leqslant\frac{1}{9}+\frac{\sqrt{30a^2+4a+2}}{24a+3}$.
From Figure 1, we see that for $0 \leqslant a \leqslant 0.482$, the upper
bound of ${\mathcal{F}\rho}$ in ($\ref{bound}$) is lower than that
given in \cite{GRJ}, i.e. the upper bound ($\ref{bound}$) is tighter than the
upper bound \cite{GRJ} in the region.

We remark that the bound obtained in Theorem 1 offers a new criterion for separability (cf. \cite{ex5}).

\section{Conclusions}
We have studied the fully entangled fraction of quantum states using the principal basis. An upper bound of FEF is given for a general bipartite state, which provides a new separability criterion. These results complement previous bounds on this subject and may give rise to new applications to the quantum information processing.

\bigskip
\noindent{\bf Acknowledgments}.
This work is supported by the NSF of China under
Grant Nos. 11401032, 11501153, 11271138, and 11531004; the NSF of Hainan Province
under Grant Nos. 20151010, 114006 and 20161006; the Scientific
Research Foundation for Colleges of Hainan Province under Grant
No. Hnky2015-18 and Simons Foundation grant 198129.

\end{document}